\documentclass [twocolumn,
 amsymb,amsmath,nobibnotes,superscriptaddress,aps,prb]{revtex4}

 \usepackage {bm}
 \usepackage{graphicx}
 \usepackage{amsmath}

\begin{document}
 \title{Tunneling through ferromagnetic barriers on the surface of a topological insulator}
 \author{Jinhua Gao}
 \affiliation{Department of Physics, and Center of Theoretical and Computational Physics, The University of Hong Kong,
 Hong Kong, China}
 \author{Wei-Qiang Chen}
 \affiliation{Department of Physics, and Center of Theoretical and Computational Physics, The University of Hong Kong,
 Hong Kong, China}
 \author{Xiao-Yong Feng}
 \affiliation{Department of Physics, and Center of Theoretical and Computational Physics, The University of Hong Kong,
 Hong Kong, China}
 \author{X. C.  Xie}
 \affiliation{Beijing National Lab for Condensed Matter Physics and Institute of Physics, Chinese Academy of Sciences,
 Beijing 100
 190, China}
 \affiliation{Department of Physics, Oklahoma State University, Stillwater, Oklahoma 74078}
 \author{Fu-Chun Zhang}
 \affiliation{Department of Physics, and Center of Theoretical and Computational Physics, The University of Hong Kong,
 Hong Kong, China}

 \date{September 2009}
 \begin{abstract}
We study the transmission  through single and double ferromagnetic barriers  on the surface of a topological insulator.
By adjusting the gate voltage and magnetization oreintation, the ferromagnetic barrier can be tuned into various
transmission regions, where the wavevector-dependent tunnelings are quite different. We find that the Klein tunneling
can be manipulated or even be turned off. These special properties  offer the possibility to control electron beams on
the ``topological metal''. Various novel devices, such as electronic collimation, wavevector filter, magnetic and
electric switchs, and wavevector-based spin valve, may be constructed based on our observed phenomena.

 \end{abstract}
 \pacs{73.43.Nq, 72.25.Dc, 85.75.-d}
 \maketitle

 \section{introduction}
Due to its unique electronic properties, the study about the graphene system has recently drawn great
attentions.\cite{ref1,ref2,ref3}. The charge carriers in  single-layer graphene behave like ``relativistic'' chiral
massless particles and possess a gapless linear energy spectrum. Effectively, in low energy regime, graphene can be
viewed as a two-dimensional (2D) Dirac fermion system. For such Dirac-like quasiparticles, the transport properties are
dramatically different from the conventional semiconductors. There is hope to utilize these peculiar transport
characteristics to develop novel relativistic electronic devices\cite{ref4}.

Recently, another 2D Dirac fermion system has been confirmed experimentally.\cite{ref5,ref6,ref7,ref8} It is called
topological metal (TM), {\it i.e.}, the surface states of a strong topological insulator
(STI)\cite{ref9,ref10,ref11,ref12,ref13,ref14}. The strong topological insulator is a new state of matter protected by
the time-reversal invariance. Though still with a bulk gap, it is distinguished from the normal insulator by the
topological $Z_2$ invariance. This distinction results in the existence of the gapless electronic states on the sample
surfaces. The surface state is the TM mentioned above, which can be effectively viewed as a 2D chiral Dirac fermion
system. Due to the time-reversal symmetry and the topology of the bulk gap, the topological surface states are robust
against disorder.

Because of the similar energy dispersions, the transport property of the TM is very similar to that of a graphene.
However, there are important differences between the two systems. TM is a surface state of semiconductor alloy, {\it e.
g.}, $Bi_{1-x}Sb_x$ or $Bi_2 Se_3$, thus more stable than a fragile thin graphene sheet. Furthermore, the spin of the
Dirac fermion of a TM is the real spin, while for graphene it is the pseudo spin. Hence, the magnetic or spin control
of the electronic transport on a TM can be much more effective than that on a graphene.
\begin {figure}
\includegraphics[width=8cm]{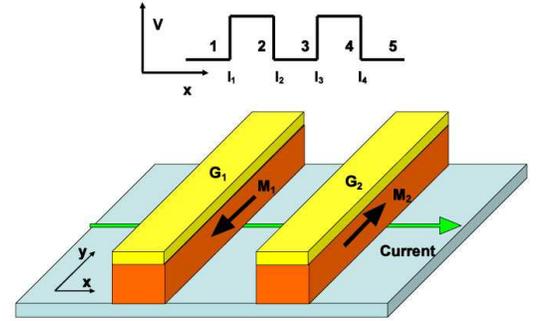}
\caption{Schematics of the ferromagnetic barriers on topological surface. Double barriers are shown as an example. The
whole system has been divided into five regions. The electrostatic potential of region $i$ is $V_i$ ($i=1\sim5$). $l_i$
denote the interface between region $i$ and $i+1$.
         }
\end  {figure}
In this work, we study the transport properties with ferromagnetic barriers on top of a topological surface. As shown
in Fig.1, for a ferromagnetic barrier, a ferromagnetic insulator is put on the top of the topological insulator to
induce an exchange field via the magnetic proximity effect. Meanwhile, a metal gate on the top of the insulator is used
to control the electrostatic potential. We find that by adjusting the magnetization orientation in the ferromagnetic
insulator and the gate voltage, the ferromagnetic barrier can be tuned into various transmission regions. The behaviors
of the transmission probability, as a function of the incident angle of the electron, in different regions are rather
different. An interesting issue is about the Klein tunneling. It is well known that, for Dirac fermion, merely with
electrostatic potential one cannot block the electron transport, and the normal incidence is at least partially
transmitted. But for the ferromagnetic barrier, because the induced exchange field, not only could the angle of Klein
tunneling be controlled, but also the Klein tunneling could even be blocked. We carefully analyze the transmission
probabilities in various regions, for both single and double barriers. Based on these analyses, we propose that
electron beam collimation, wavevector filtering, magnetic and electric switch, wavevector-based in-plane spin valve
could be realized in such structures. These transport phenomena are quite different from the case with ferromagnetic
barrier on graphene\cite{ref15, ref16, ref17,ref18}.

This paper is organized as follows. In Sec. II, we introduce the model and formalisms. In Sec. III, we discuss the
analytical and numerical results of the transport properties of single and double ferromagnetic barriers. And we then
show how to construct novel electronic devices based on the phenomena obtained in our theory. Finally, a brief summary
is given in Sec. IV.

\section{model and formalism}
The Dirac fermion of the topological surface can be described as:
\begin{equation}
  H=\hbar v_f \textbf{k} \cdot \mathbf{\sigma}  + \textbf{M} \cdot   \mathbf{\sigma}
\end{equation}
where $\textbf{M}= M(\sin\theta_m \cos \phi_m, \sin\theta_m \sin\phi_m, \cos\theta_m )$ is the effective exchange field
induced via magnetic proximity effect. In this work, we only consider the in-plane exchange field. The main consequence
of the z component of the exchange field is to open an energy gap and its effects have been investigated in a recent
theoretical study\cite{ref19}.
 If the Fermi level of
the whole system is  inside the gap, the transport of the barrier is completely blocked. No tunneling is permitted.
Otherwise, the effect is just a change of the position of the Fermi surface. But in our ferromagnetic barrier model,
adjusting the gate voltage will also change the Fermi surface. So for simplicity and without loss of generality, we
merely consider  an in-plane exchange field. The sketch of the barrier structure is shown in Fig.1.

Take the single barrier structure for example, the whole system consists of three different regions and the interfaces
between these regions are labeled by $l_i$. With the above Hamiltonian, the eigenfunctions for region $\alpha$ are
\begin{equation}
 \phi^{(1)}_{\alpha}(x,y)= e^{i k^+_{\alpha x} \cdot x + k_{\alpha y}  \cdot y }
 \left(\begin{array}{cccc}
1 \\ S_\alpha e^{i \theta _ \alpha} \\
\end{array}\right)
\end{equation}
\begin{equation}
 \phi^{(2)}_{\alpha}(x,y)= e^{i k^-_{\alpha x} \cdot x + k_{\alpha y}  \cdot y }
 \left(\begin{array}{cccc}
1 \\ S_\alpha e^{i (\pi - \theta _ \alpha)} \\
\end{array}\right)
\end{equation}
where $S_\alpha = \epsilon - V_\alpha$, $e^{i \theta_\alpha}= (k_{\alpha x} + i k_{\alpha y})/ |\textbf k_\alpha|$, and
$k^\pm_{\alpha x}= -M \cos\phi_{m\alpha}  \pm \sqrt{(\epsilon - V_\alpha)^2 -  (k_{\alpha y}+M \sin\phi_{m\alpha})^2}$.
$V_\alpha$ is the potential of this region, $\epsilon$ is the energy of the incident electron. Due to the translation
invariant along the y-axis, the momentum $k_y$ is conserved, i.e. $k_{\alpha y}= k_y$.

In order to solve the transport problem, we use the standard transfer matrix method\cite{ref20}.  By matching the wave
functions at the interface $l_1=0$, we get the transfer matrix
\begin{equation}
\begin{split}
T_{l_1}&=T^{(0)}_{21}= \frac{1}{2S_2 \cos\theta_2} \\& \cdot \left( \begin{array}{cccc}
                                               S_2 e^{-i \theta_2} + S_1 e^{i \theta_1} & S_2 e^{-i \theta_2} - S_1
                                               e^{i \theta_1} \\
                                                S_2 e^{i \theta_2} - S_1 e^{i \theta_1}& S_2 e^{i \theta_2} + S_1 e^{i
                                                \theta_1}
                                              \end{array}\right)
                                              \end{split}
\end{equation}
The transfer matrix at the interface $l_2 = a$  can be represented as
\begin{equation}
\begin{split}
T_{l_2}= &\left( \begin{array}{cccc}
                 e^{-ik^+_{3x} a}  &  0 \\
                 0               &  e^{-ik^-_{3x} a}
                 \end{array}
         \right)
         \cdot T^{(0)}_{32} \\& \cdot
         \left( \begin{array}{cccc}
                 e^{ik^+_{2x} a}  &  0 \\
                 0               &  e^{ik^-_{2x} a}
                 \end{array}
         \right)
         \end{split}
\end{equation}
where $T^{(0)}_{32}$ is the same as $T^{(0)}_{21}$ except the region index. Finally, we will get the total transfer
matrix for the single barrier
\begin{equation}
T_s=T_{l_2}T_{ l_1}
\end{equation}
With the transfer matrix, the reflection and translation coefficients can be got through following equation
\begin{equation}
\left( \begin{array}{cccc}
        t \\
        0
       \end{array}
\right) = T_s \cdot \left( \begin{array}{cccc}
        1 \\
        r
       \end{array}
\right)
\end{equation}
 $T=|t|^2$ is just the transmission probability we want.

Double barrier structure could be treated in  similar way. The only difference is that we have four interfaces here.
\begin{equation}
  T_d = T_{l_4} T_{l_3}T_{l_2}T_{l_1}
\end{equation}

With the transmission probability, in the linear transport regime and for low temperature, we can get the conductance
density (the conductance per unit length along the interface) of the barrier structure\cite{ref19}
\begin{equation}
g=g_0 \int^{\pi /2}_{-\pi /2} d \theta_{in} T(\epsilon_f, \theta_{in}) \cos\theta_{in}
\end{equation}
Here, $g_0=\frac{e^2 \epsilon_f }{4 \pi^2 \hbar^2 v_f}$, $\epsilon_f$ is the Fermi energy of the system, $\theta_{in}$
is the incident angle of the electron and $T(\epsilon_f, \theta_{in})$ is the transmission probability through the
barrier.

\section{Result and discussion}
\subsection{Single barrier}
For a single barrier, an analytical expression of the transmission probability could be given.
\begin{widetext}
\begin{equation}
T_s=|t|^2 =\frac{\cos^2 \theta_1  \cos ^2 \theta_2}{\cos ^2 \theta_1 \cos^2 \theta_2 \cos^2 (q_xa) + \sin^2 (q_x a)
(1-S_1 S_2 \sin \theta_1 \sin \theta_2)^2}
\end{equation}
\end{widetext}
where $q_x = k^+_{2x}+M\cos \phi_{m}$. Formally, it is the same as the one for the graphene potential barrier except a
shift of the Fermi surface due to the in-plane exchange field.

\begin {figure}
\includegraphics[width=8cm]{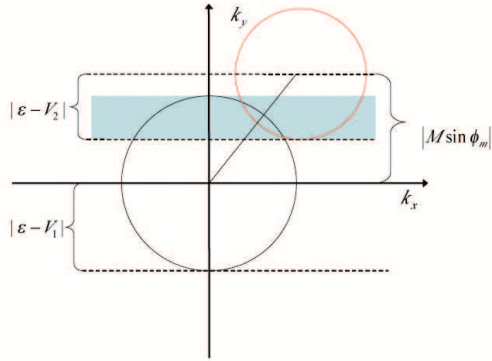}
\caption{(Color online) The overlap of the Fermi surfaces of region 1 and 2. Here, region 2 is a ferromagnetic barrier
and its fermi surface (red circle) has been shifted due to the induced exchange field $M$. $\epsilon$ is the energy of
incident electron. $V_{1(2)}$ is the electrostatic potential of region 1(2). $\phi_m$ is the in-plane rotational angle
of the exchange field. }
\end  {figure}

As mentioned above, for a ferromagnetic barrier, changing  the gate voltage will alter the Fermi surface, the in-plane
exchange field will also cause a shift.  The key issue about the transport is that, during the tunneling,  the energy
and $k_y$ of the incident electron are conserved. As shown in Fig. 2, in the incident region (Region 1 in Fig.1), the
range of $k_y$ values $|\epsilon - V_1|>|k_y|$ is plotted. But not all the incident electrons are allowed to transmit,
since the Fermi surface has been changed in the ferromagnetic barrier ( Region 2 in Fig. 1). The $k_y$ range of the
ferromagnetic barrier is $|\epsilon - V_2|>|k_y + M \sin \phi_m|$. Therefore, the permitted incident angle of the
electron is determined by the overlap of the $k_y$ ranges of adjacent regions. It means  that, through adjusting the
gate voltage and the direction of the magnetization of the ferromagnetic barrier, we can tune the acceptable range of
$k_y$, and rather different transmissions will appear accordingly.

\begin {figure}
\includegraphics[width=9cm]{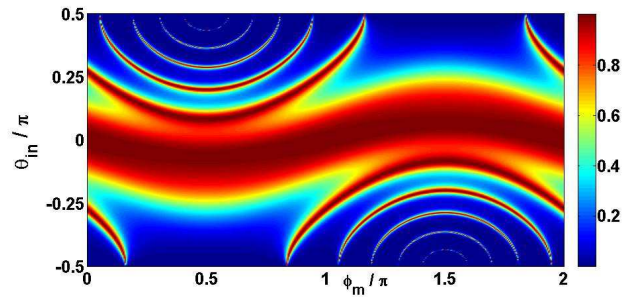}\\
\caption{(Color online) The transmission probability of the single ferromagnetic barrier in the total transmission
status. $\theta_{in}$ is the incident angle of the electron. $\phi_m$ is the in-plane rotation angle of the
magnetization of the barrier. Here, $v_f = 10^6 ms^{-1}$ , $\epsilon = 80 meV$, $V_1 = 0$, $V_2 = 200 meV$, $M = 40
meV$, $l_1 = 0$, $l_2 = 100  nm$.
         }
\end  {figure}
\begin {figure}
\includegraphics[width=9cm]{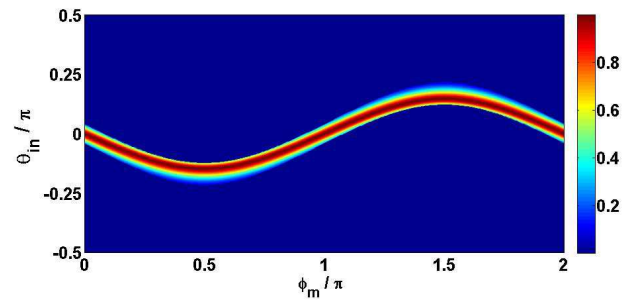}\\
\caption{(Color online) The transmission probability of the single ferromagnetic barrier in the partial transmission
status. $V_2=90 meV$ and other parameters are the same as Fig. 3.
         }
\end  {figure}

\begin {figure}
\begin{tabular}{c}
\includegraphics[width=8cm]{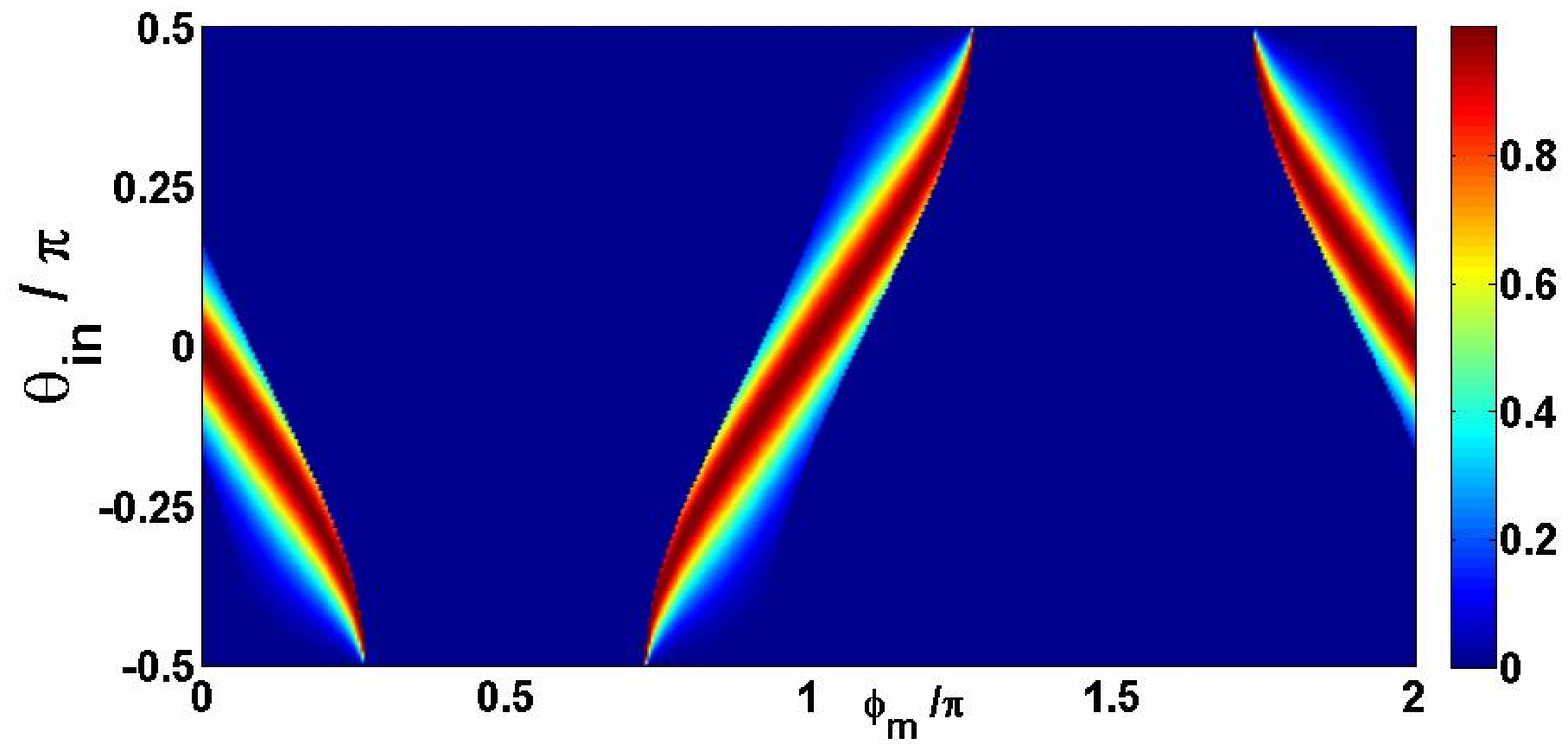} \\
\includegraphics[width=8cm]{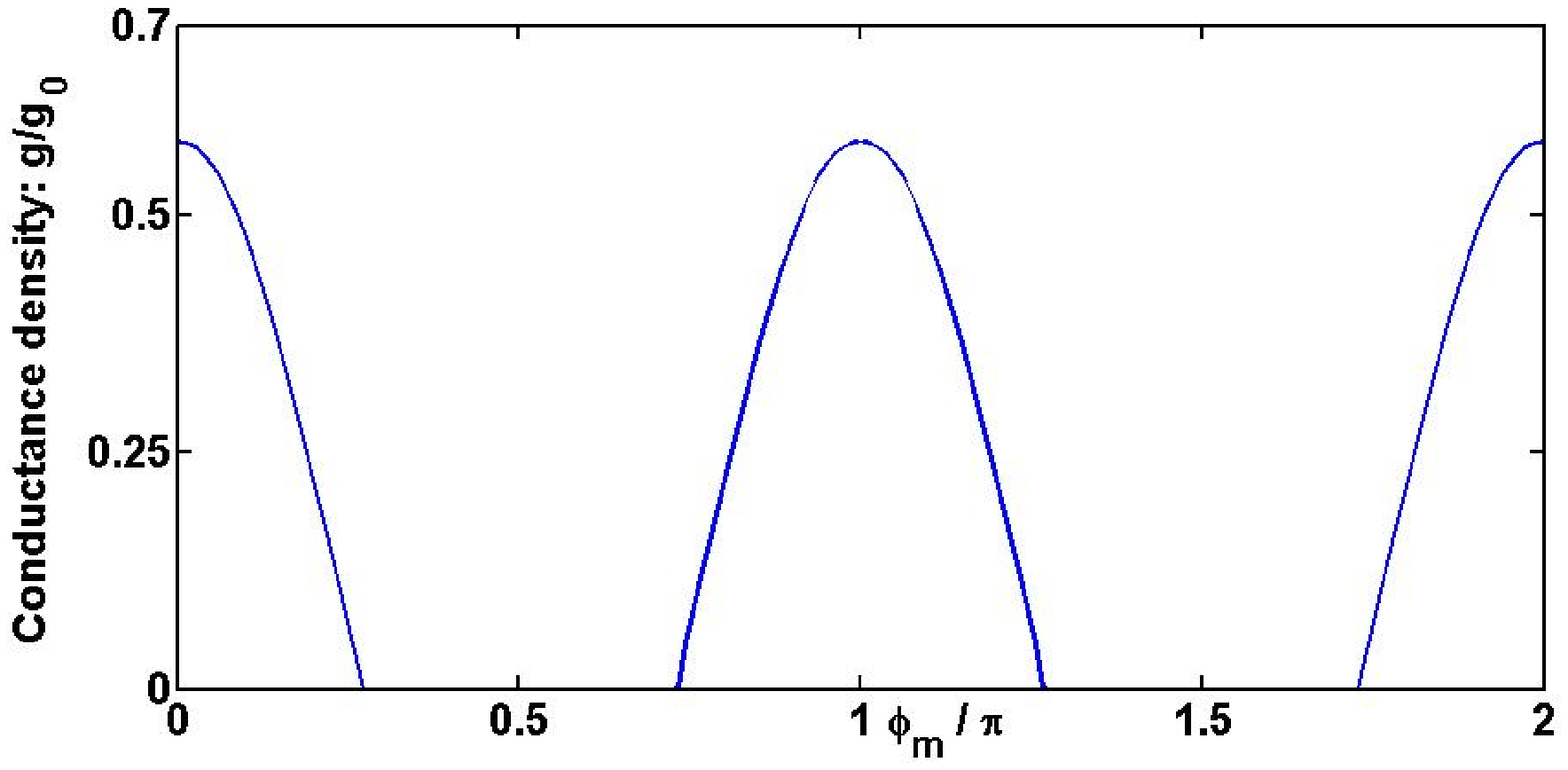}
\end{tabular}
\caption{(Color online) Up:The transmission probability of the single ferromagnetic barrier in the blockable status.
$V_2=90  meV$ , $V_1 = 60 meV$ and other parameters are the same as Fig. 3. Down: The corresponding conductance density
in this case.
         }
\end  {figure}

\begin {figure}
\begin{tabular}{c}
\includegraphics[width=8cm]{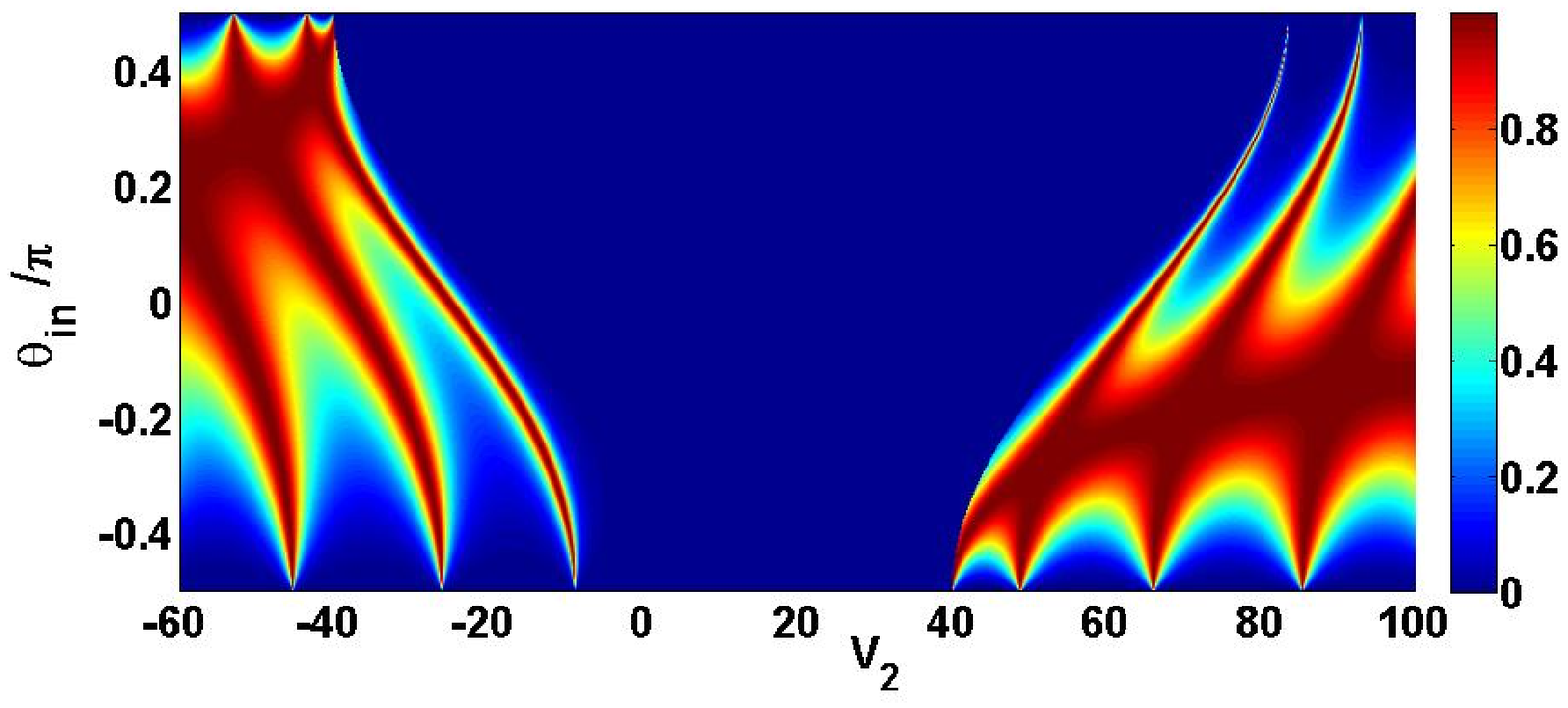} \\
\includegraphics[width=8cm]{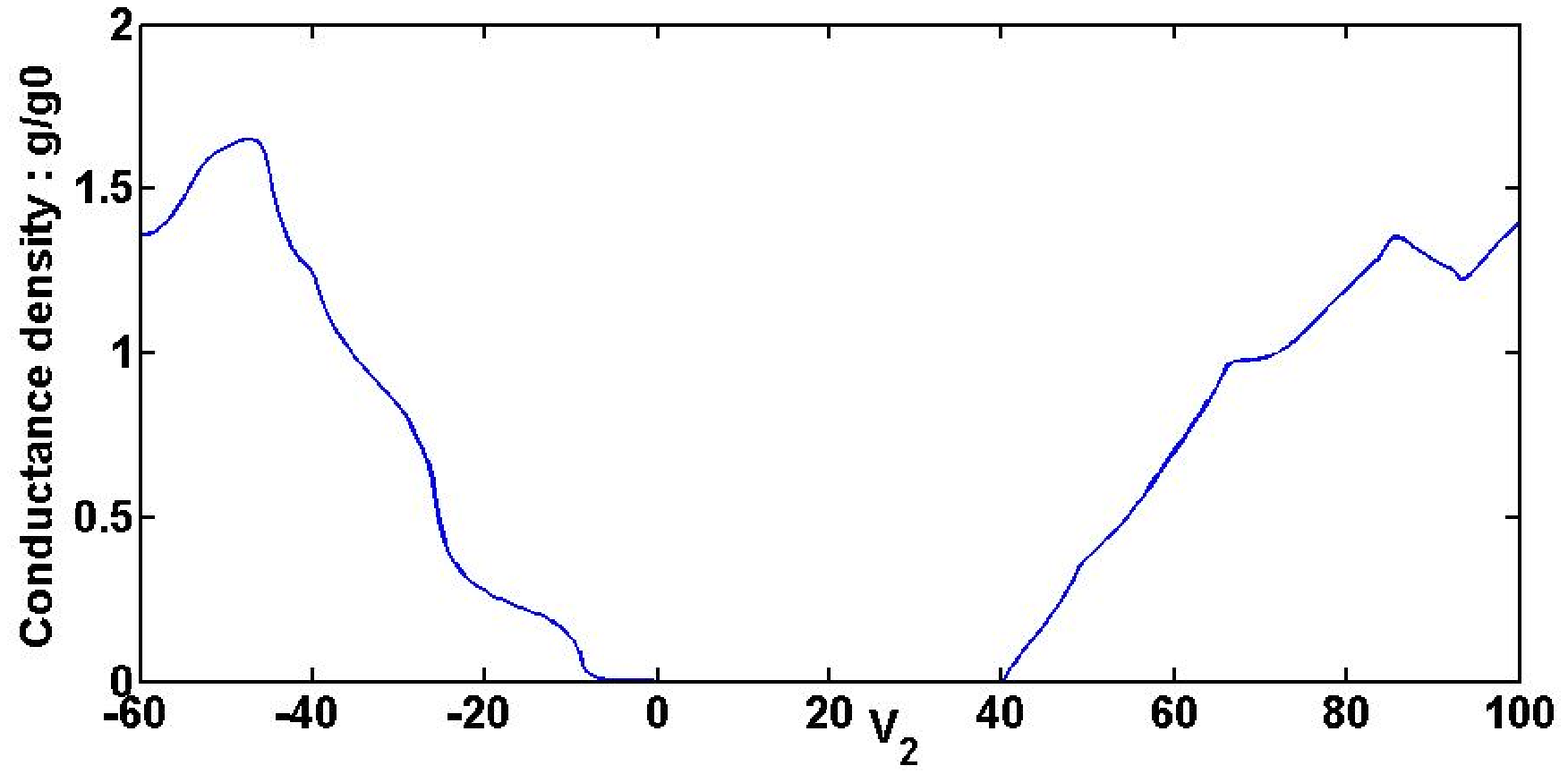}
\end{tabular}
\caption{(Color online) Up:The transmission probability of the single ferromagnetic barrier in the blockable status.
$\phi_m = 0.5 \pi$ and other parameters are the same as Fig. 5.
 Down: The corresponding conductance density in this case.
         }
\end  {figure}

In the case of Fig. 3, $V_2$ is far from the energy of incident electron $\epsilon$.  So for any in-plane rotation
angle $\phi_m$, there are always  $|\epsilon - V_2| - M\sin \phi_m > |\epsilon - V_1|$ and $-|\epsilon - V_1| > -
|\epsilon - V_2| -M \sin \phi_m$. Hence, all the incident angles are allowed from the viewpoint of $k_y$ conservation.
We can say that the ferromagnetic barrier is in the total transmission status now. As shown clearly, when we rotate the magnetization orientation, the angle of the Klein tunneling will be changed and new resonance peaks will appear.

Another intriguing case occurs when the gate voltage of the barrier $V_2$ is near the energy of the incident electron
$\epsilon$. In this situation, the relations,  $|\epsilon - V_1| > |\epsilon - V_2| - M\sin \phi_m $ and  $ - |\epsilon
- V_2| - M \sin \phi_m
>-|\epsilon - V_1|$ are always hold for any angle $\phi_m$ . As
shown in Fig. 4,  we can get a transmission sector of incident angle in this partial transmission status. The width of
the sector is determined by $|\epsilon - V_2|$ and its position can be tuned through adjusting the magnetization
orientation. It obviously can be used as a wavevector filter. Furthermore, as pointed in electron optics, this has
immediate applicability in electron beam collimation and lensing\cite{ref21, ref22}. Better collimation effect could be
achieved through a series of barriers.

When $-M - |\epsilon - V_2|> |\epsilon - V_1|$, for some angle $\phi_m$, the overlap between the $k_y$ ranges of the
two regions will be absent and the conductance becomes zero. This is the blockade status, in which the transmission of
the ferromagnetic barrier can be completely blocked. Since that both $V_2$ and $\phi_m$ could be used to tune the
overlap,  we can construct magnetic and electric on-off with the single ferromagnetic barrier structure. Fig. 5 gives
the case of magnetic switch. With a definite $V_2$, there is a blocked sector of angle $\phi_m$, in which the
transmission probability is zero for any incident electron. The conductance of the magnetic switch is shown as well.
The alteration of electrostatic potential can also completely block the transmission. This is quite different from a
normal tunneling barrier for Dirac fermions. It is known, due to the Klein paradox, merely with electrostatic potential
one can not block the transport of 2D Dirac fermions. However, with a ferromagnetic barrier the condition is different.
For a definite angle $\phi_m$, if $|M \sin \phi_m| > |\epsilon - V_1|$, there must be a critical gate voltage $V_{2c}$
that $-M \sin \phi_m- |\epsilon - V_{2c}| = |\epsilon -V_1|$. Compared with $V_{2c}$, when $|\epsilon - V_2|$
decreases, the $k_y$ ranges of the two regions separate out and the transport is turned off (See in Fig. 6). It is
just an electric switch which the on/off state of the current is controlled by the gate voltage. Interestingly, when
increasing $V_2$, the current will be turned on (See in Fig. 6). This unique property, i.e. the voltage window for the
open status, is not possible with a normal semiconductor barrier. Clearly, this transport phenomenon could be used to
distinguish TM from the normal semiconductor and the graphene systems. It means that transport measurement is also a
very useful tool to experimentally confirm the existence of the TM.

\subsection{Double barrier}
\begin {figure}
\begin{tabular}{c}
\includegraphics[width=8cm]{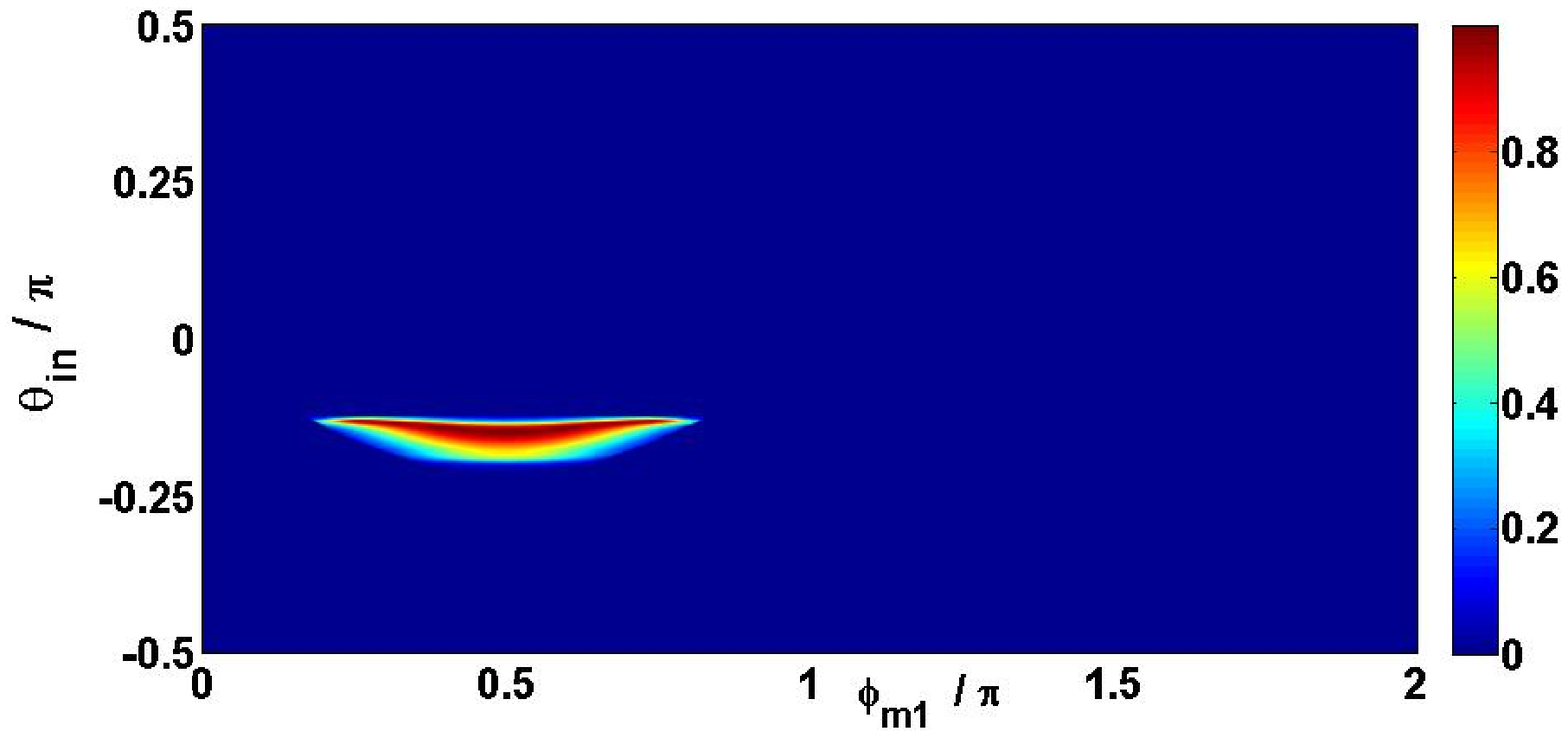} \\
\includegraphics[width=8cm]{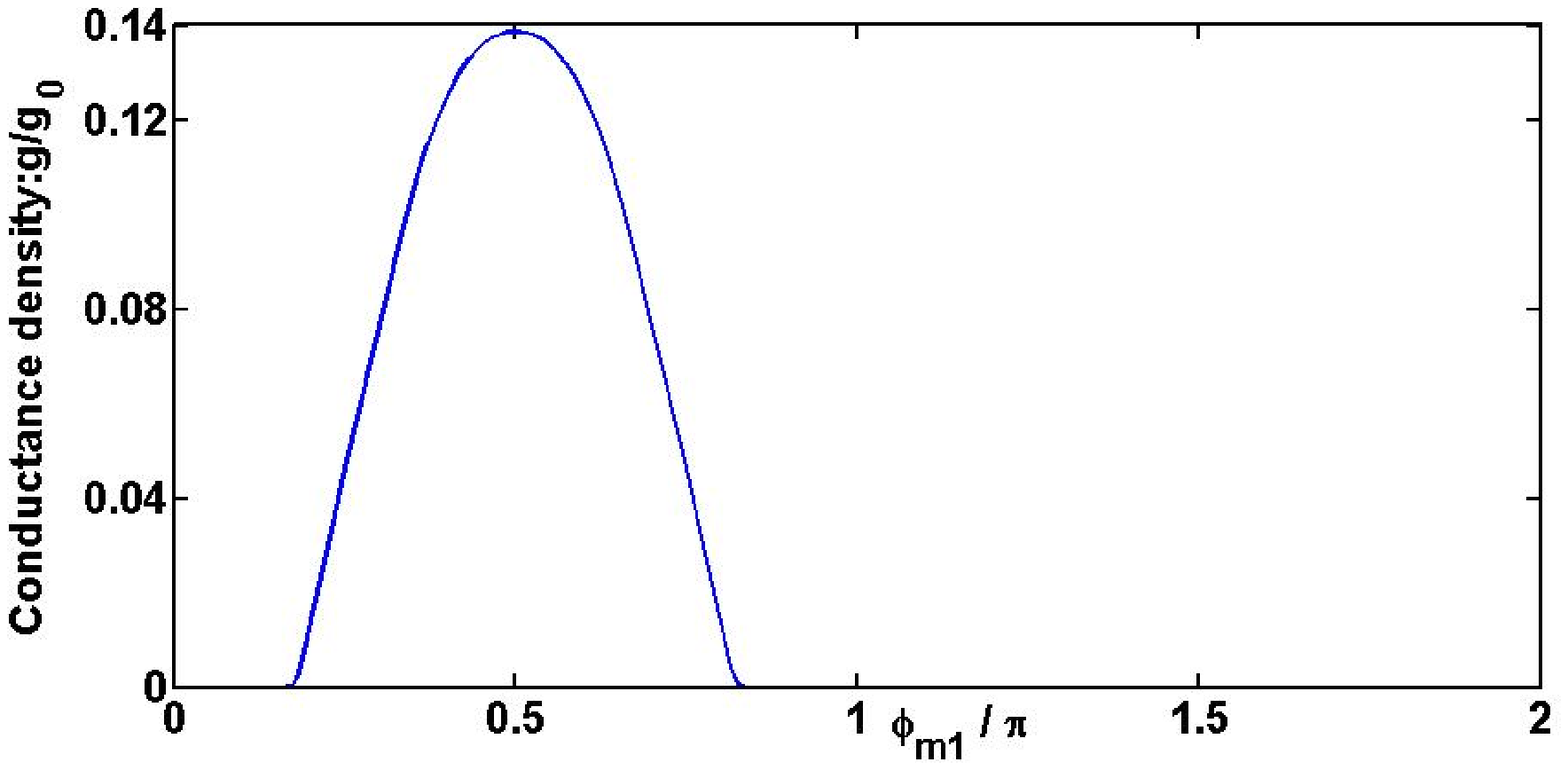}
\end{tabular}
\caption{(Color online) Up:The transmission probability of the double ferromagnetic barrier structure. $\phi_{m1(2)}$
is the in-plane rotation angle of ferromagnetic barrier 1(2), and  $\phi_{m2}= 0.5 \pi$. Both barriers are in the
partial transmission status. $V_1 = V_3 = V_5 = 60meV$, $V_2 = V_4 =90 meV$,  $\epsilon =80 meV$ and $l_2 -l_1 = l_3-
l_2 = l_4 - l_3 = 100nm$. Down: The corresponding conductance density in this case
         }
\end  {figure}

\begin {figure}
\includegraphics[width=8cm]{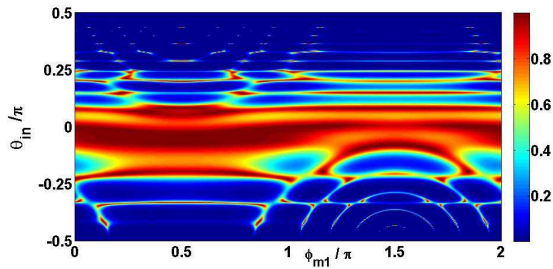}
\caption{(Color online) The transmission probability of the double ferromagnetic barrier structure. Both barriers are
in the total transmission status. $V_1 = V_3 = V_5 = 0$, $V_2 = V_4 =200 meV$, $\epsilon =80 meV$ and $l_2 -l_1 = l_3-
l_2 = l_4 - l_3 = 100nm$.
         }
\end  {figure}
For the double barrier structure, We also could understand the transport properties through the conservation of the
$k_y$ and the energy. As mentioned above, depending on the voltage and magnetization, single ferromagnetic barrier
could be tuned into three different transmission statuses: total transmission , partial transmission , and the blockade
situations. The transport behavior of the double barrier system is more complex than that of the single one, since that
the two barriers can be tuned into different transport situations.

Here, we investigate two interesting examples. One is the case that two ferromagnetic barriers are both in the partial
transmission status. In this case, both barriers have their own transmission sector of the incident angle of the
electron. Only when the two transmission sectors overlap, the double barrier structure becomes conducting. Hence, as
shown in Fig. 7, the direction of magnetization of the second barrier is fixed $\phi_{m2} = \pi / 2$. Only if the angle
of the first barrier $\phi _{m1}$ is tuned in parallel, the transmission of the double barrier structure is nonzero.
The corresponding conductance  is also given in Fig. 7. Essentially, in this situation the double barrier system
becomes a special wavevector-dependent in-plane spin valve system, for which the parallel configuration is the only
conducting state. The blockade of the transmission is mainly because the mismatch of the wavevectors, while for  normal
spin valve the primary reason is the different spin dependent potentials. Therefore, for a normal spin valve, the
conductance will vary continuously with the angle between the magnetizations. But, as shown in Fig. 7, for the in-plane
spin valve on TM the change of the conductance is rather sharp and the the conductance of the non-parallel
configuration is exactly zero. These special transport properties of the double barrier structure on TM could also be
tested in transport experiments as a distinction between the TM and normal semiconductors.

For a single barrier, it is hard to directly observe the partial transmission region through the conductance. This is
because that the change of  position of the transmission sector will not induce an obvious change in the conductance.
But in the double barrier system this can be done.

As a comparison (See in Fig. 8), we also calculate the case that both barriers are in the total transmission status. In
this situation, the angle dependent transmission probability is much more complicated.

\section{Summary}
In conclusion, we have investigate the transport properties of the ferromagnetic barriers on the surface of a
topological insulator. The single and double barrier structures have been carefully considered. We show that the gate
voltage and the magnetization could be used to manipulate the Fermi surface of the barrier. Due to the conservation of
energy and $k_y$ of an incident electron during the transport process, adjusting the Fermi surface will tune the single
ferromagnetic barrier into three different transmission statuses: total transmission , partial transmission , and the
transmission blockade. In the total transmission case, all the incident angles are allowed. The angle of the Klein
tunneling and position of the resonance vary with the direction of the magnetization, and meanwhile new resonance peaks
appear. In the partial transmission situation, only the incident angles in a transmission sector are allowed. The
position and width of the transmission sector can be adjusted through the gate voltage and the magnetization.
Therefore, the electron collimation and wavevector filter could be realized based on this phenomenon. With proper gate voltage and
magnetization, one can even separate the Fermi surfaces, reaching the blockade status. In this case, no transport can
occur, even the Klein tunneling. We propose that both magnetic and electric switchs, based on the single ferromagnetic
barrier on topological surface, could be realized in this situation. We also study the transport of the double barrier
structure. In this system, we can construct a special wavevector-dependent spin valve when both barriers are in the
partial transmission status. The spin valve is conducting only if the magnetization of the two barriers are parallel.
We emphasis that these unique transport properties of TM could be used to distinguish TM from the normal semiconductors
and the graphene systems. Hence, the observations of these transport properties provide confirmation of the existence
of TM.

\textbf{Acknowledgments:} We  wish to acknowledge the partial support from RGC grant of HKSAR. XCX is supported by DOE
with grant number DE-FG02-04ER46124 and C-Spin center of Oklahoma.

 \end{document}